\documentclass[a4paper,11pt]{article}

\usepackage{setup}
\usepackage{jheppub}
\usepackage{todonotes}
\usepackage{placeins}
\usepackage{comment}

\usepackage{cite}
\usepackage{ifthen}
\usepackage{subcaption}
\usepackage{siunitx}
\usepackage{enumitem}
\usepackage{booktabs}
\usepackage{bookmark}

\usepackage{cleveref}
\DeclareSIUnit\fb{\femto\barn}

\renewcommand{\tabcolsep}{1em}

\def\lapprox{\lower .7ex\hbox{$\;\stackrel{\textstyle <}{\sim}\;$}}
\def\gapprox{\lower .7ex\hbox{$\;\stackrel{\textstyle >}{\sim}\;$}}
\definecolor{lightgray}{HTML}{A6A39A}
\definecolor{darkgray}{HTML}{504E48}
\definecolor{silver}{HTML}{E0DFDE}
\definecolor{brown}{HTML}{5F4541}
\definecolor{beige}{HTML}{DCCCAC}
\definecolor{green}{HTML}{345F53}
\definecolor{yellow}{HTML}{F6B65A}
\definecolor{blue}{HTML}{568BCF}
\definecolor{red}{HTML}{AE1932}
\definecolor{orange}{HTML}{D16F15}

\makeatletter
\newcommand{\myitem}[1]{%
	\item[#1]\protected@edef\@currentlabel{#1}%
}
\makeatother

\preprint{{\raggedleft%
ZU-TH 59/23 
}}

\title{Four-jet event shapes in hadronic Higgs decays}
\author[a,b]{Aude Gehrmann--De Ridder,}
\author[a,c]{Christian T~Preuss,}
\author[d]{Ciaran Williams}

\affiliation[a]{Institute for Theoretical Physics, ETH, CH-8093 Z\"urich, Switzerland}
\affiliation[b]{Department of Physics, University of Z\"urich, CH-8057 Z\"urich, Switzerland}
\affiliation[c]{Department of Physics, University of Wuppertal, 42119 Wuppertal, Germany}
\affiliation[d]{Department of Physics, University at Buffalo, The State University of New York, Buffalo 14260, U.S.A.}

\emailAdd{gehra@phys.ethz.ch}
\emailAdd{preuss@uni-wuppertal.de}
\emailAdd{ciaranwi@buffalo.edu}

\abstract{
  We present next-to-leading order perturbative QCD predictions for four-jet-like event-shape observables in hadronic Higgs decays.
  To this end, we take into account two Higgs-decay categories: involving either the Yukawa-induced decay to a $\Pqb\Paqb$ pair or the loop-induced decay to two gluons via an effective Higgs-gluon-gluon coupling.
  We present results for distributions related to the event-shape variables thrust minor, light-hemisphere mass, narrow jet broadening, $D$-parameter, and Durham four-to-three-jet transition variable. For each of these observables we study the impact of higher-order corrections and compare their size and shape in the two Higgs-decay categories.
  We find large \NLO corrections with a visible shape difference between the two decay modes, leading to a significant shift of the peak in distributions related to the $\PH\to\Pg\Pg$ decay mode.  
}

\begin{document}
\maketitle
\flushbottom

\section{Introduction}
\label{sec:intro}

Future lepton colliders, such as the FCC-ee \cite{FCC:2018evy}, the CEPC \cite{CEPCStudyGroup:2018ghi}, or the ILC \cite{ILC:2013jhg}, are projected to operate as so-called ``Higgs factories'', producing an unprecedented amount of events which contain a Higgs boson in the final state.
In leptonic collisions, Higgs bosons are predominantly produced via the Higgsstrahlung process $\Pep\Pem\to \PZ\PH$ at low centre-of-mass energies ($\lesssim 450~\GeV$) and via the vector-boson-fusion process $\Pep\Pem\to \Pl \Pal \PH$ at high centre-of-mass energies ($\gtrsim 450~\GeV$).
The clean experimental environment of leptonic collisions, in which contamination from QCD backgrounds such as initial-state radiation and multi-parton interactions is absent, will allow for precise measurements of Higgs-boson properties, such as its branching ratios and total width.
In particular, it will become possible to have access to so-far unobserved subleading hadronic decay channels such as the decay to gluons or $\Pqc$-quark pairs, which are currently inaccessible in hadron-collider environments, where only the dominant $\PH\to\Pqb\Paqb$ decay was observed to date \cite{ATLAS:2018kot,CMS:2018nsn}.

A well suited class of observables used to probe the structure of QCD radiation are so-called event shapes. Event shapes provide direct access to the geometric properties of hadronic events, while at the same time being amenable to perturbative calculations.
As such, they have played a vital role in precision studies of the hadronic $\upgamma^*/\PZ \to \Pq\Paq$ decay at LEP, enabling for instance precise determinations of the strong coupling constant \cite{Dissertori:2007xa,Dissertori:2009ik,Bethke:2009ehn,Hoang:2015hka,Verbytskyi:2019zhh,Kardos:2020igb}.
It is customary to divide event shapes into three- and four-jet event shapes. We define four-jet event shapes as those observables that are non-zero for topologies with four resolved partons and vanish in the limit of planar three-particle configurations\footnote{
We note that this differs from the nomenclature sometimes used in resummed calculations, where these are referred to as three-jet event shapes, representing the number of hard radiators.}.
Historically, experimental measurements at LEP have mostly focused on three-jet event shapes in hadronic $\upgamma^*/\PZ$ decays, matched by a plethora of theoretical results, such as next-to-next-to-leading order (\NNLO) in QCD corrections \cite{Gehrmann-DeRidder:2007vsv,Gehrmann-DeRidder:2009fgd,Gehrmann-DeRidder:2014hxk,Weinzierl:2009nz,Weinzierl:2009ms,Weinzierl:2009yz,DelDuca:2016ily,Kardos:2018kth}, resummation of logarithmically enhanced contributions \cite{Becher:2011pf,Becher:2012qc,Balsiger:2019tne,Hoang:2014wka,Banfi:2014sua,Bhattacharya:2022dtm,Bhattacharya:2023qet}, and hadronisation effects \cite{Gehrmann:2010uax,Luisoni:2020efy,Caola:2021kzt,Caola:2022vea,Agarwal:2023fdk}.
Four-jet event shapes, in $Z$ decays on the other hand, have had less attention, despite the fact that they give access to important properties of the strong interactions, such as the quadratic Casimirs \cite{Kluth:2003yz}.
Theoretical calculations of four-jet event shapes have generally achieved \NLO accuracy in $\Pep\Pem$ collisions \cite{Signer:1996bf,Dixon:1997th,Parisi:1978eg,Nagy:1997mf,Nagy:1997yn,Nagy:1998bb,Nagy:1998kw,Campbell:1998nn}. For most of these, also at least the next-to-leading logarithmic (\NLL) contributions are known \cite{Banfi:2000si,Banfi:2000ut,Banfi:2001sp,Banfi:2001pb,Larkoski:2018cke,Arpino:2019ozn}.

In hadronic Higgs decays, three-jet-like event shapes such as the thrust and energy-energy correlators have been discussed as discriminators of fermionic and gluonic decay channels \cite{Gao:2016jcm,Gao:2019mlt,Luo:2019nig,Gao:2020vyx}.
In \cite{Coloretti:2022jcl}, the full set of the six ``classical'' three-jet-like event-shape observables related to the event-shape variables thrust, $C$-parameter, heavy-hemisphere mass, narrow and wide jet broadening, and the Durham three-to-two-jet transition variable  (called $y_{23}$) have been calculated at \NLO QCD for hadronic Higgs decays. The results obtained for the two Higgs-decay categories were compared. It was shown that for all event shapes the size of \NLO corrections were larger in the $\PH\Pg\Pg$ category. The distributions related to the three-jet event-shape variable thrust and $y_{23}$ showed the most striking shape differences between Higgs decay categories. It was therefore argued that these event shapes could act as discriminators between the two Higgs-decay categories highlighting where the $\PH\Pg\Pg$ category could be enhanced.   
Recently, a novel method to determine branching ratios in hadronic Higgs decays via fractional energy correlators has been proposed in \cite{Knobbe:2023njd}.
To the best of our knowledge, four-jet event shapes have not yet been computed in hadronic Higgs decay processes. 

With this paper, we aim at delivering, for the first time, theoretical predictions for four-jet-like event shapes in hadronic Higgs decays including perturbative QCD corrections up to the \NLO level.
The event-shape distributions computed here are related to the event-shape variables narrow jet broadening, $D$-parameter, light-hemisphere mass, thrust minor, and the four-to-three-jet transition variable in the Durham algorithm (called $y_{34}$).
It is the purpose of this work to compare the impact of higher-order corrections in terms of size, shape, and perturbative stability of these four-jet-like event-shape distributions.

Furthermore, besides its direct applicability in Higgs phenomenology at lepton colliders, our calculation provides necessary ingredients needed for the \NNLO and \NthreeLO calculations of jet observables in three-jet and two-jet final states of hadronic Higgs decays, which are currently only known for decays to $\Pqb$-quarks \cite{Mondini:2019gid,Mondini:2019vub}.

The paper is structured as follows. In \cref{sec:higgs_decays}, we discuss the ingredients of our computation for both Higgs-decay categories up to \NLO QCD.
Our predictions for the five event-shape distributions are presented in \cref{sec:results}.
We summarise our findings and give an outlook on future work in \cref{sec:conclusion}. 

\begin{figure}
    \centering
    \includegraphics{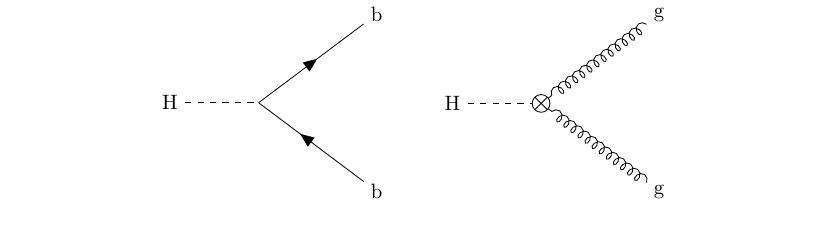}
    \caption{Feynman diagrams of the two decay categories of a Higgs boson decaying into two jets, $\PH\to\Pqb\Paqb$ (\textit{left}) and $\PH\to\Pg\Pg$ (\textit{right}).}
    \label{fig:h2j_diags}
\end{figure}

\section{Hadronic Higgs decays up to NLO QCD}
\label{sec:higgs_decays}

Hadronic Higgs decays proceed via two classes of processes; either involving the decay to a quark-antiquark pair, $\PH \to \Pq \Paq$, or the decay to two gluons, $\PH \to \Pg\Pg$. 
Within the scope of this work, we consider a five-flavour scheme and assume all light quarks, including the $\Pqb$-quark, to be massless. In order to allow the Higgs boson to decay to a $\Pqb\Paqb$ pair, we keep a non-vanishing Yukawa coupling for the $\Pqb$-quark. 
The coupling between gluons and the Higgs is considered in an effective theory in the limit of an infinitely heavy top-quark.
Based on this, we will classify parton-level processes induced by the two Born-level decay modes into two categories.

\begin{figure}[t]
    \centering
    \includegraphics{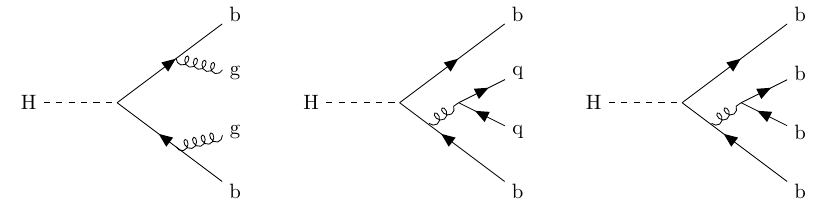}
    \caption{Tree-level Feynman diagrams of four-parton processes in the $\PH\to\Pqb\Paqb$ decay category contributing to our calculation at \LO. Here, $\Pq$ denote any flavour $\Pqu,\Pqd,\Pqs,\Pqc,\Pqb$.}
    \label{fig:diagsHbb4j}
\end{figure}

In the first class of processes, the Higgs decays to a $\Pqb$-quark pair, mediated by the Yukawa coupling $y_\Pqb$ between the Higgs and the bottom quark, and the decay is computed using the Standard-Model Lagrangian. The two-parton decay diagram associated to this decay mode is shown in the left-hand panel of \cref{fig:h2j_diags}. In the remainder of this paper, we shall refer to these decay modes as belonging to the $\PH\Pqb\Paqb$ category. 
The relevant four- and five-parton processes contributing to this decay category, entering our calculation at \LO and \NLO, respectively, are shown in the first column of \cref{tab:channels}. Representative Feynman diagrams for the tree-level four-parton processes contributing at \LO are shown in \cref{fig:diagsHbb4j}.

\begin{figure}[t]
    \centering
    \includegraphics{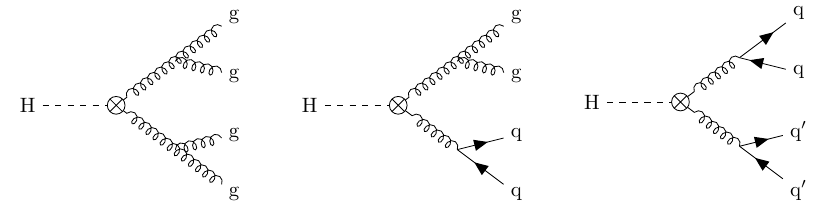}
    \caption{Tree-level Feynman diagrams of four-parton processes in the $\PH\to\Pg\Pg$ decay category contributing to our calculation at \LO. Here, $\Pq$ and $\Pq'$ denote any flavour $\Pqu,\Pqd,\Pqs,\Pqc,\Pqb$, possibly with $\Pq' = \Pq$ but $\Pq' \neq \Pq$ in general.}
    \label{fig:diagsHgg4j}
\end{figure}

In the second class of processes, the decay proceeds via a top-quark loop. In the limit of an infinitely large top-quark mass, where the top quarks decouple, we compute these decays in an effective theory with a direct coupling of the Higgs field to the gluon field-strength tensor.
In this second category, which we shall refer to as the $\PH\Pg\Pg$ category, the interaction is mediated by an effective $\PH\Pg\Pg$ vertex, which is represented as a crossed dot in the two-parton decay diagram shown in the right-hand part of \cref{fig:h2j_diags}.
A summary of the four- and five-parton processes contributing to this decay category, entering our computation at \LO and \NLO, are presented in the second column of \cref{tab:channels}. Feynman diagrams for the tree-level four-parton processes contributing at \LO are shown in \cref{fig:diagsHgg4j}.

Following the nomenclature presented in \cite{Coloretti:2022jcl}, the preceding discussion can be cast into an effective Lagrangian containing both decay categories as,
\begin{equation}
    \mathcal{L}_\text{Higgs} = -\frac{\lambda(M_\Pqt,\muR)}{4}HG^a_{\mu\nu}G^{a,\mu\nu} + \frac{y_\Pqb(\muR)}{\sqrt{2}}H\bar\psi_\Pqb\psi_\Pqb \, .
\label{eq:lagrangian}
\end{equation}
Here, the effective Higgs-gluon-gluon coupling proportional to $\alphas$ is given by
\begin{equation}
    \lambda(M_\Pqt,\muR) = -\frac{\alphas(\muR)C(M_\Pqt,\muR)}{3\uppi v} \, ,
\end{equation}
where we define the Higgs vacuum expectation value $v$ and the top-quark Wilson coefficient $C(M_\Pqt,\muR)$. 
The $\Pqb$-quark Yukawa coupling on the other hand is given by 
\begin{equation}
    y_\Pqb(\muR) = \frac{4\uppi\alpha}{\sqrt{2}M_\PW\sin\theta_\text{W}} \overline{m}_\Pqb(\muR) \, ,
\label{eq:YukawaCoupling}
\end{equation}
and depends directly on the $\Pqb$-quark mass.
As indicated by the renormalisation-scale dependence, both, the effective $\PH\Pg\Pg$ coupling and the $\PH\Pqb\Paqb$ Yukawa coupling, are subject to renormalisation, which we here perform in the $\overline{\text{MS}}$ scheme with $\NF = 5$.
While the top-quark Wilson coefficient $C(M_\Pqt,\muR)$ is known up to $\order{\alphas^4}$ in the literature \cite{Inami:1982xt,Djouadi:1991tk,Chetyrkin:1997iv,Chetyrkin:1997un,Chetyrkin:2005ia,Schroder:2005hy,Baikov:2016tgj}
we here only need to consider the first-order expansion of the Wilson coefficient $C(M_\Pqt,\muR)$ in $\alphas$. It is given by
\begin{equation}
    C(M_\Pqt,\muR) = 1 + \frac{11}{6}\NC\frac{\alphas(\muR)}{2\uppi} + \order{\alphas^2} \, .
\label{eq:WilsonCoefficient}
\end{equation}
which is independent of the top-quark mass $M_\Pqt$.
Further, the Yukawa mass $\bar{m}_{\Pqb}$ runs with $\muR$, with the  
the running of the Yukawa coupling $y_{\Pqb}$ with $\muR$ taken into account via the results of \cite{Vermaseren:1997fq}.

Finally, to conclude some general remarks, we wish to highlight the importance for the present study that the kinematical mass of the $\Pqb$-quark be vanishing.
For kinematically massless $\Pqb$-quarks, the operators in the effective Lagrangian given in \cref{eq:lagrangian} do not interfere or mix under renormalisation, cf.~e.g.~\cite{Gao:2019mlt}. In particular, this means that the product of Yukawa-induced amplitudes with HEFT-induced amplitudes vanishes to all orders in $\alphas$ when computing squared matrix elements. Moreover, it implies that the Wilson coefficient $C(M_\Pqt,\muR)$ in \cref{eq:WilsonCoefficient} and the Yukawa coupling $y_\Pqb(\muR)$ in \cref{eq:YukawaCoupling} evolve independently under the renormalisation group.
As a result, it is possible to consider two distinct decay categories, $\PH\to\Pqb\Paqb$ and $\PH\to\Pg\Pg$, and calculate theoretical predictions and higher-order corrections for each of them separately, as done previously in \cite{Coloretti:2022jcl} for the case of three-jet-like event shapes.
The effect of interferences between the Yukawa-induced decay $\PH\to \Pqb\Paqb$ and the HEFT-induced $\PH\to\Pg\Pg$ to order $\order{\alphas^3}$ in a framework with massive $\Pqb$-quarks and $\Pqc$-quarks has been studied in \cite{Mondini:2020uyy}.

\begin{table}[t]
  \centering
  \caption{Partonic channels contributing to the decay $\PH \to 4j$ at \LO and \NLO.}
  \begin{tabular}{llll}\toprule
    ~    & $\PH\to\Pqb\Paqb$ type & $\PH\to\Pg\Pg$ type & ~\\ \midrule
    \LO  & ~ & $\PH \to \Pg \Pg \Pg \Pg$ & tree-level \\
    ~    & $\PH \to \Pqb \Paqb \Pg \Pg$ & $\PH \to \Pq \Paq \Pg \Pg$ & tree-level \\
    ~    & $\PH \to \Pqb \Paqb \Pq \Paq$ & $\PH \to \Pq \Paq \Pq' \Paq'$ & tree-level \\
    ~    & $\PH \to \Pqb \Paqb \Pqb \Paqb$ & $\PH \to \Pq \Paq \Pq \Paq$ & tree-level \\ \midrule
    \NLO & ~ & $\PH \to \Pg \Pg \Pg \Pg$ & one-loop \\
    ~    & $\PH \to \Pqb \Paqb \Pg \Pg$ & $\PH \to \Pq \Paq \Pg \Pg$ & one-loop \\
    ~    & $\PH \to \Pqb \Paqb \Pq \Paq$ & $\PH \to \Pq \Paq \Pq' \Paq'$ & one-loop \\
    ~    & $\PH \to \Pqb \Paqb \Pqb \Paqb$ & $\PH \to \Pq \Paq \Pq \Paq$ & one-loop \\
    ~    & ~ & $\PH \to \Pg \Pg \Pg \Pg \Pg$ & tree-level \\
    ~    & $\PH \to \Pqb \Paqb \Pg \Pg \Pg$ & $\PH \to \Pq \Paq \Pg \Pg \Pg$ & tree-level \\
    ~    & $\PH \to \Pqb \Paqb \Pq \Paq \Pg$ & $\PH \to \Pq \Paq \Pq' \Paq' \Pg$ & tree-level \\
    ~    & $\PH \to \Pqb \Paqb \Pqb \Paqb \Pg$ & $\PH \to \Pq \Paq \Pq \Paq \Pg$ & tree-level \\ \bottomrule
  \end{tabular}
  \label{tab:channels}
\end{table}
In the remainder of this section we discuss in detail the ingredients of our computation. We split the discussion in the following parts. In \cref{subsec:framework}, we present the general framework enabling us to compute Higgs decay observables up to NLO, while 
in \cref{subsec:Hbb} and \cref{subsec:Hgg} we discuss details of the ingredients and their numerical implementation in the two Higgs-decay categories.
In \cref{subsec:validation}, we provide a summary of the checks performed to validate our results.

\subsection{General framework}
\label{subsec:framework}
Given an infrared-safe observable $O$, the differential four-jet decay rate of a colour-singlet resonance of mass $M$, like the Standard-Model scalar Higgs boson, can be written at each perturbative order up to \NLO level (i.e., including corrections up to the third order in the strong coupling $\alphas$) as
\begin{equation}
    \frac{1}{\Gamma^{(n)}(\muR)}\frac{\rd\Gamma}{\rd O}(\muR,O) = \left(\frac{\alphas(\muR^2)}{2\uppi}\right)^2 \frac{\rd \bar B}{\rd O} + \left(\frac{\alphas(\muR^2)}{2\uppi}\right)^3 \left(\frac{\rd \bar C}{\rd O} + \beta_0\log\left(\frac{\muR^2}{M^2}\right)\frac{\rd \bar B}{\rd O} \right) \, ,
\label{eq:rate}
\end{equation}
where $\rd \bar B$ and $\rd \bar C$ denote the differential \LO and \NLO coefficients, respectively. 
Subject to the order of the calculation, $n$, the differential decay rate in \cref{eq:rate} is normalised to the \LO ($n=0$) or \NLO ($n=1$) partial width, $\Gamma^{(1)}$ or $\Gamma^{(0)}$, respectively.

Schematically, the \LO coefficient $B$ can be determined as
\begin{equation}
    \left(\frac{\alphas}{2\uppi}\right)^2\frac{\rd \bar B}{\rd O} = \frac{1}{\Gamma^{(n)}} \int \frac{\rd\Gamma^\text{B}}{\rd \Phi_4}\delta(O-O(\Phi_4))\, \rd\Phi_4 \, ,
\end{equation}
while the \NLO coefficient C is obtained as
\begin{multline}
    \left(\frac{\alphas}{2\uppi}\right)^3\frac{\rd \bar C}{\rd O} = \frac{1}{\Gamma^{(n)}}\int \left[\frac{\rd \Gamma^\text{V}}{\rd \Phi_4} + \frac{\rd \Gamma^\text{T}_\text{NLO}}{\rd \Phi_4}\right] \delta(O-O(\Phi_4))\, \rd\Phi_4 \\
    + \frac{1}{\Gamma^{(n)}}\int \left[\frac{\rd \Gamma^\text{R}}{\rd \Phi_5}\delta(O-O(\Phi_5)) - \frac{\rd \Gamma^\text{S}_\text{NLO}}{\rd \Phi_5}\delta(O-O(\Phi_4(\Phi_5)))\right]\, \rd\Phi_5 \, .
\end{multline}
Here, $\rd\Gamma^\text{R}$ and $\rd\Gamma^\text{V}$ denote the real and virtual (one-loop) correction differential in the four-parton and five-parton phase space, respectively. The real and virtual subtraction terms $\rd\Gamma^\text{S}_\NLO$ and $\rd\Gamma^\text{T}_\NLO$, on the other hand, ensure that the real and virtual corrections are separately infrared finite and make them amenable to numerical integration. The notation $\Phi_4(\Phi_5)$ represents a kinematic mapping from the five-parton to the four-parton phase space, specific to the subtraction term $\rd \Gamma^\text{S}_\NLO$.
The choice of subtraction terms is in principle arbitrary and subject only to the requirements that the virtual subtraction term $\rd \Gamma^\text{T}_\NLO$ cancels all explicit poles in $\rd\Gamma^\text{V}$, the real subtraction term $\rd\Gamma^\text{S}_\NLO$ cancels all implicit singularities in $\rd\Gamma^\text{R}$, and the net contribution of the subtraction terms to the decay width vanishes,
\begin{equation}
    \int\frac{\rd\Gamma^\text{T}_\NLO}{\rd\Phi_4}\,\rd\Phi_4 = \int\frac{\rd\Gamma^\text{S}_\NLO}{\rd\Phi_5}\,\rd\Phi_5 \, .
\end{equation}
The last requirement is equivalent to demanding that $\rd\Gamma^\text{T}_\NLO$ be the integral of $\rd\Gamma^\text{S}_\NLO$ over the respective one-particle branching phase space.
To compute our predictions for the four-jet like hadronic event shapes in Higgs decays, 
we here rely on the antenna-subtraction framework \cite{Campbell:1998nn,Gehrmann-DeRidder:2005btv,Currie:2013vh} to construct all subtraction terms and implement our calculation in the publicly available\footnote{\url{http://eerad3.hepforge.org}} \eerad framework \cite{Gehrmann-DeRidder:2014hxk}.
This code has previously been used to study event shapes \cite{Gehrmann-DeRidder:2007vsv,Gehrmann-DeRidder:2009fgd} and jet distributions \cite{Gehrmann-DeRidder:2008qsl} in $\Pep\Pem \to 3j$ at \NNLO and was recently extended to include hadronic Higgs decays to three jets at \NLO \cite{Coloretti:2022jcl}.
Our new implementation builds upon the latter and is done in a flexible manner, utilising the existing infrastructure, such as phase-space generators of the  $\Pep\Pem\to 3j$ calculation at \NNLO.
In particular, the real-radiation contributions proportional to $\alphas^2$ of the three-jet computation of Higgs-decay observables enter our current calculation of four-jet-like observables at the Born level.
We implement all matrix elements and subtraction terms in analytic form, enabling a fast and numerically stable evaluation of the perturbative coefficients up to \NLO level.

\subsection{Yukawa-induced contributions}
\label{subsec:Hbb}
Four-particle tree-level matrix elements in the $\PH\Pqb\Paqb$ category are taken from the \NLO $\PH\to\Pqb\Paqb j$ process in \cite{Coloretti:2022jcl}, which have been calculated explicitly using \form \cite{Kuipers:2012rf}. Tree-level five-parton matrix elements are taken from the \NthreeLO $\PH \to  \Pqb\Paqb$ and \NNLO $\PH \to \Pqb\Paqb j$ calculation in \cite{Mondini:2019gid,Mondini:2019vub}, in turn calculated using BCFW recursion relations \cite{Britto:2005fq}.
Similarly, one-loop four-parton amplitudes are taken from the same calculation \cite{Mondini:2019gid,Mondini:2019vub}. These have been derived analytically by use of the generalised unitarity approach \cite{Bern:1994zx}, using quadruple cuts for box coefficients \cite{Britto:2004nc}, triple cuts for triangle coefficients \cite{Forde:2007mi}, double cuts for bubble coefficients \cite{Mastrolia:2009dr}, and $d$-dimensional unitarity techniques for the rational pieces \cite{Badger:2008cm}.

As per \cref{eq:rate}, the differential four-jet rate depends on the partial two-parton width. At \NLO, the inclusive width of the $\PH\to\Pqb\Paqb$ decay reads
\begin{equation}
  \Gamma_{\PH\to\Pqb\Paqb}^{(1)}(\muR) = \left[1 + \left(\frac{\alphas(\muR^2)}{2\uppi}\right)\left(\frac{17}{2}\CF + 3\CF\log\left(\frac{\muR^2}{M_\PH^2}\right) \right) \right]\Gamma_{\PH\to\Pqb\Paqb}^{(0)}(\muR) \, ,
  \label{eq:widthHbbNLO}
\end{equation}
where the LO partial width is given by
\begin{equation}
    \Gamma_{\PH\to\Pqb\Paqb}^{(0)}(\muR) = \frac{y_{\Pqb}^2(\muR)M_\PH\NC}{8\uppi} \, .
\end{equation}
The running of the Yukawa coupling $y_{\Pqb}$ with $\muR$ present in the partial width is taken into account via the results of \cite{Vermaseren:1997fq}.

\subsection{Effective-theory contributions}
\label{subsec:Hgg}
In the $\PH\Pg\Pg$ category, we take the four-parton tree-level matrix elements from the \NLO $\PH\to 3j$ calculation in the $\PH\Pg\Pg$ category presented in \cite{Coloretti:2022jcl}.
Five-parton tree-level and four-parton one-loop amplitudes are obtained by crossing the ones used in the $pp \to \PH+1j$ \NNLO calculation in \nnlojet \cite{Chen:2014gva,Chen:2016zka}. The five-parton tree-level amplitudes are based on the results presented in \cite{Campbell:2010cz,DelDuca:2004wt,Badger:2004ty,Dixon:2004za}, while the four-parton one-loop amplitudes are based on the results presented in \cite{Campbell:2010cz,Dixon:2004za,Ellis:2005qe,Badger:2006us,Badger:2007si,Glover:2008ffa,Badger:2009hw,Dixon:2009uk,Badger:2009vh}.
The four-parton decay rate further receives contributions from the $\order{\alphas}$ expansion of the top-quark Wilson coefficient $C(M_\Pqt,\muR)$. We implement these as a finite contribution to the virtual correction,
\begin{equation}
    \rd\Gamma^\text{V} \to \rd\Gamma^\text{V} + \left(\frac{\alphas}{2\uppi}\right)\frac{11}{3}\NC\rd\Gamma^\text{B} \, .
\end{equation}

As per \cref{eq:rate}, the differential four-jet rate depends on the partial two-parton width. At NLO, the inclusive width of the $\PH\to\Pg\Pg$ decay reads
\begin{equation}
  \Gamma_{\PH\to\Pg\Pg}^{(1)}(\muR) = \left[1 + \left(\frac{\alphas(\muR^2)}{2\uppi}\right)\left(\frac{95}{6}\CA - \frac{7}{3}\NF + 2\beta_0\log\left(\frac{\muR^2}{M_\PH^2}\right)\right) \right] \Gamma_{\PH\to\Pg\Pg}^{(0)}(\muR) \, ,
  \label{eq:widthHggNLO}
\end{equation}
where the LO partial width is given by
\begin{equation}
    \Gamma_{\PH\to\Pg\Pg}^{(0)}(\muR) = \frac{\alphas^2(\muR)G_\text{F}M_\PH^3}{36\uppi^2\sqrt{2}} \, .
\end{equation}

\subsection{Validation}
\label{subsec:validation}
We have performed numerical checks of all matrix elements on a point-by-point basis against automated tools.
Tree-level four- and five-parton matrix elements are tested against results generated with \madgraph \cite{Alwall:2011uj,Alwall:2014hca}, whereas four-parton one-loop matrix elements are validated using \openloops2 \cite{Buccioni:2019sur}.
In all cases, we have found excellent agreement between our analytic expressions and the auto-generated results.

Real subtraction terms $\rd \Gamma^\mathrm{S}_\NLO$ have been tested numerically using so-called ``spike tests'', first introduced in the context of the antenna-subtraction framework in \cite{Pires:2010jv} and applied to the \NNLO di-jet calculation in \cite{NigelGlover:2010kwr}.
To this end, trajectories into singular limits are generated using \text{Rambo} \cite{Kleiss:1985gy} and the ratio $\rd \Gamma^\mathrm{S}_\NLO/\rd \Gamma^\mathrm{R}$ is evaluated on a point-by-point basis. Where applicable, azimuthal correlations are included via summation over antipodal points.
In all relevant single-unresolved limits, we have found very good agreement between the subtraction terms and real-radiation matrix elements.

To confirm the cancellation of explicit poles in the virtual corrections $\rd\Gamma^\mathrm{V}$ by the virtual subtraction terms $\rd\Gamma^\mathrm{T}_{\NLO}$, we have both checked the cancellation analytically and numerically.

\section{Results}
\label{sec:results}

In the following, we shall present numerical results for the five different four-jet event shapes: thrust minor, light-hemisphere mass, narrow jet broadening, $D$-parameter, and the four-to-three-jet transition variable in the Durham algorithm.
We discuss our numerical setup and scale-variation prescription in \cref{subsec:setup}, before defining the four-jet event-shape observables in \cref{subsec:observables}. Theoretical predictions are shown and discussed in \cref{subsec:predictions}.

\subsection{Numerical setup and scale-variation prescription}
\label{subsec:setup}
We consider on-shell Higgs decays with a Higgs mass of $M_\PH = 125.09~\GeV$. We work in the $G_{\upmu}$-scheme and consider electroweak quantities as constant parameters. Specifically, we consider the following electroweak input parameters
\begin{equation}
  G_{\mathrm{F}} = 1.20495 \times 10^{-5}~\GeV^{-2} \, , \quad M_\PZ = 91.1876~\GeV \, , \quad M_\PW = 80.385~\GeV \, ,
\end{equation}
yielding a corresponding value of $\alpha = \frac{1}{128}$.
As alluded to above, we keep a vanishing kinematical mass of the $\Pqb$-quark throughout the calculation, but consider a non-vanishing Yukawa mass. The running of the Yukawa mass with $\muR$ is calculated using the results of \cite{Vermaseren:1997fq}, corresponding to $\bar{m}_\Pqb(M_\PH)$ close to $2.61~\GeV$.

We choose the Higgs mass as renormalisation scale, $\muR = M_\PH$, and apply a scale variation $\muR \to k_\mu \muR$ about this central scale with $k_\mu \in \left[\frac{1}{2},2\right]$. We wish to point out that this prescription also affects the normalisation of our distributions via \cref{eq:rate}.
We use one- and two-loop running for the strong coupling $\alphas$, at \LO and \NLO respectively, obtained by solving the renormalisation-group equation at the given order, as detailed in \cite{Gehrmann-DeRidder:2007vsv,Gehrmann-DeRidder:2014hxk}.
For the strong coupling, we choose a nominal value at scale $M_\PZ$ given by 
\begin{equation}
    \alphas(M_\PZ) = 0.1179 \, ,
\end{equation}
corresponding to the current world average \cite{ParticleDataGroup:2022pth}.

\subsection{Four-jet event-shape observables}
\label{subsec:observables}
We consider the five different four-jet event-shape observables related to the following event shapes thrust minor, light-hemisphere mass, narrow jet broadening, $D$-parameter, and the Durham four-to-three-jet transition variable. They are defined as follows \cite{Campbell:1998nn}:

\paragraph{Thrust minor}
We define thrust minor as
\begin{equation}
    T = \left(\frac{\sum\limits_j \vert \vec p_j \cdot \vec n_\text{Min} \vert }{\sum\limits_k \vert\vec p_k\vert}\right) \, ,
\end{equation}
where $\vec n_\text{Min}$ is given by $\vec n_\text{Min} = \vec n_\text{T} \times \vec n_\text{Maj}$.
Here, $\vec n_\text{T}$ is the thrust axis \cite{Brandt:1964sa,Farhi:1977sg}, defined as the unit vector which maximises
\begin{equation}
    T = \max_{\vec n}\left(\frac{\sum\limits_j \vert \vec p_j \cdot \vec n \vert }{\sum\limits_k \vert\vec p_k\vert}\right) \, ,
\end{equation}
and $\vec n_\text{Maj}$ is a unit vector for which in addition $\vec n_\text{Maj} \cdot \vec n_\text{T} = 0$ holds. 
The thrust minor measures the distribution of particles orthogonal to the plane defined by $\vec n_\text{T}$ and $\vec n_\text{Maj}$. It vanishes for topologies with less than four particles, as three particles always span a plane parallel to the thrust and thrust-major axis and two particles exactly align with the thrust axis.

\paragraph{Light-hemisphere mass}
Starting from the thrust axis $\vec n_\text{T}$, events are divided into two hemispheres (jets), $H_1$ and $H_2$, separated by an axis orthogonal to $\vec n_\text{T}$. For each hemisphere, the hemisphere mass is calculated as
\begin{equation}
    \frac{M_i^2}{s} = \frac{1}{E_\text{vis}^2} \left(\sum\limits_{j \in H_1} p_j\right)^2 \, ,
\end{equation}
with $E_\text{vis}$ the visible energy in the event.
The light hemisphere mass is then given by the smaller of the two hemisphere masses \cite{Clavelli:1979md},
\begin{equation}
    \frac{M_\text{L}^2}{s} = \min\limits_{i\in \{1,2\}}\left(\frac{M_i^2}{s}\right) \, .
\end{equation}
For massless particles, the light-hemisphere mass is non-vanishing only for topologies with at least four particles, as the lighter hemisphere will otherwise always consist of a single particle.

\paragraph{Narrow jet broadening} 
Jet broadenings measure the distribution of the transverse momenta of particles with respect to the thrust axis. The narrow jet broadening is defined as the jet-broadening value of the more narrow hemisphere,
\begin{equation}
    B_\text{N} = \min(B_1,B_2) \, ,
\end{equation}
with the hemisphere broadenings $B_i$ given by \cite{Catani:1992jc}
\begin{equation}
    B_i = \frac{\sum\limits_{j\in H_i}\vert \vec p_j \times \vec {\color{red}n_\text{T}} \vert}{2\sum\limits_k \vert \vec p_k\vert} \, ,
\end{equation}
for the two hemispheres $H_1$ and $H_2$ defined by the thrust axis.
Since a single particle has a vanishing hemisphere broadening, the narrow jet broadening is zero for events with less than four particles, for which always one of the hemispheres contains only a single particle.

\paragraph{$D$-parameter}
The $D$-parameter measures the planarity of an event and vanishes for planar configurations. As such, it is non-zero only for events containing at least four particles, as three (or less) particles always span a plane. It is defined as \cite{Parisi:1978eg}
\begin{equation}
    D = 27 \lambda_1\lambda_2\lambda_3 \, ,
\end{equation}
given in terms of the three eigenvalues $\lambda_1$, $\lambda_2$, and $\lambda_3$ of the linearised momentum tensor,
\begin{equation}
    \Theta^{\alpha\beta} = \frac{1}{\sum\limits_k \vert\vec p_k\vert} \sum\limits_j \frac{p^\alpha_jp^\beta_j}{\vert\vec p_j\vert} \, , \text{ with } \alpha,\beta \in \{1,2,3\} \, .
\end{equation}

\paragraph{Four-to-three-jet transition variable}
The four-to-three-jet transition variable denoted by $y_{34}$ corresponds to the jet resolution parameter $y_\mathrm{cut}$ at which an event changes from a four-jet to a three-jet event according a specific clustering algorithm. 
Here, we specifically consider the Durham algorithm, in which the distance measure reads
\begin{equation}
    y_{ij}^\text{D} = \frac{2E_iE_j(1-\cos\theta_{ij})}{E_\text{vis}^2} \, .
\end{equation}
We use the so-called E-scheme, in which the four-momenta of the two particles are added linearly in each step of the algorithm, 
\begin{equation}
    p_{ij} = p_i + p_j \, .
\end{equation}

\subsection{Infrared behaviour}
\label{subsec:infraredBehaviour}
All observables introduced in \cref{subsec:observables} are non-vanishing only for at least four resolved particles. In the region of phase space where only three jets can be resolved, i.e., where the observables of \cref{subsec:observables} become small, the differential rate becomes enhanced by large logarithms of the form $\log^m(1/O)$. In this phase-space region, fixed-order calculations become unreliable and a faithful calculation of the differential decay width requires the resummation of the large logarithms.

To avoid large logarithmic contributions spoiling the convergence of our fixed-order calculation in the three-jet region, we restrict the range of validity of our predictions as follows:
we impose a small cut-off $y_0 = 10^{-5}$ on linear distributions and $y_0 = \re^{-7}$ on logarithmically binned distributions.
We consider this minimal value for all observables mentioned above in \cref{subsec:observables} 
with the exception of the distribution related to the four-to-three-jet transition variable $y_{34}$, where we require a cut off at $y_0 = \re^{-10}$. The $y_0$ cut-offs are imposed on both four- and five-parton configurations and ensure the reliability of our predictions in the whole kinematical region considered. 

In addition to the observable cutoff $y_0$, we introduce a technical cut-off $t_\mathrm{cut}$ and require that all dimensionless two-parton invariants in the five-parton states to be above this cut, $s_{ij}/s > t_\mathrm{cut}$. This technical cut-off parameter improves the numerical stability of the antenna-subtraction procedure by avoiding large numerical cancellations between the real subtraction term and the real matrix element.
By default, we employ a technical cut-off of $t_\mathrm{cut} = 10^{-7}$.

We note that there is a subtle interplay of the observable cut-off $y_0$ and the technical cut-off $t_\mathrm{cut}$. We have studied this interplay for all observables considered in \cref{subsec:observables} by varying the technical cut-off between $t_\mathrm{cut} \in \{10^{-8}, 10^{-7}, 10^{-6}\}$ and verified that this variation leaves all distributions above the observable cut-off $y_0$ unaffected.
We further wish to emphasise that the independence on $t_\mathrm{cut}$ also validates the correct implementation of our subtraction terms.

\subsection{Comparison of predictions in both Higgs-decay categories}
\label{subsec:predictions}
In this subsection, we present theoretical predictions at \LO and \NLO QCD for the event shapes defined in \cref{subsec:observables}. For each event shape, we show two binnings in the observable; a linear binning to highlight the general structure and a logarithmic binning to emphasise the behaviour of the distribution in the infrared region, in particular the position of its peak.
In all cases, we present results according to \cref{eq:rate}. In particular, we normalise by the \LO (\NLO) partial two-jet decay width for \LO (\NLO) distributions.\footnote{We refrain from reweighting by the branching ratios of the $\PH\Pqb\Paqb$ and $\PH\Pg\Pg$ decays as done in \cite{Coloretti:2022jcl}.}

\begin{figure}
    \centering
    \includegraphics[width=0.45\textwidth]{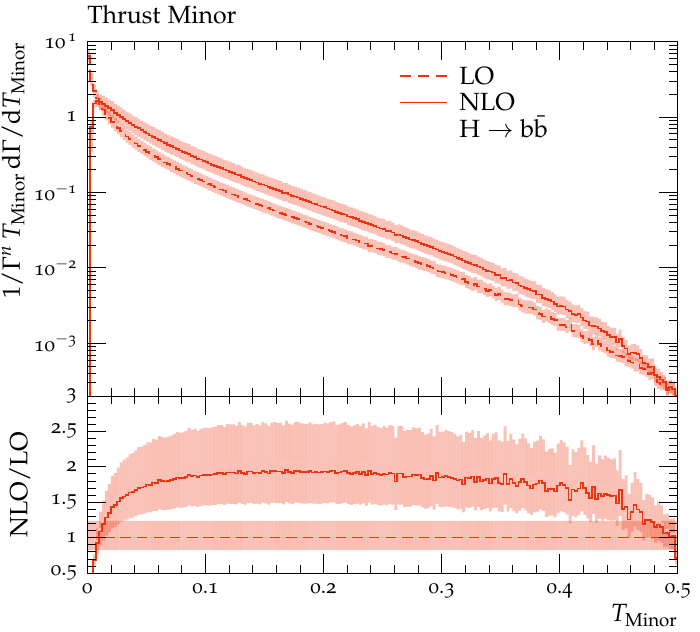}
    \includegraphics[width=0.45\textwidth]{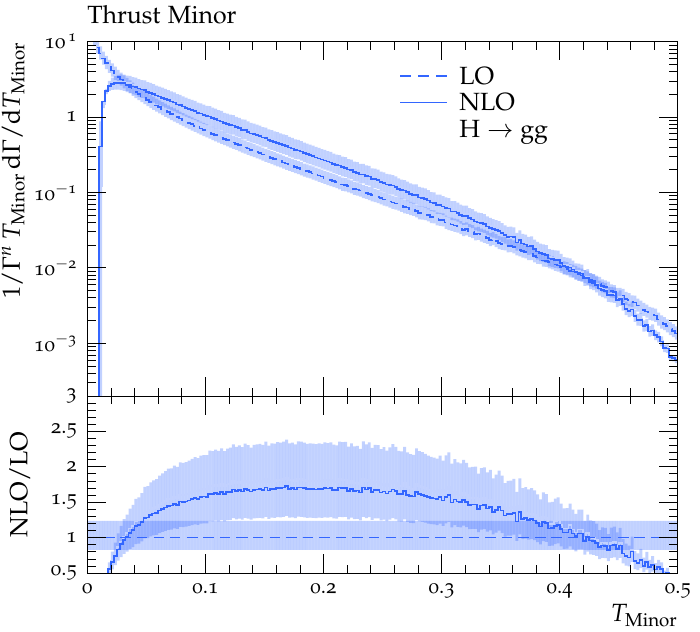} \\
    \includegraphics[width=0.45\textwidth]{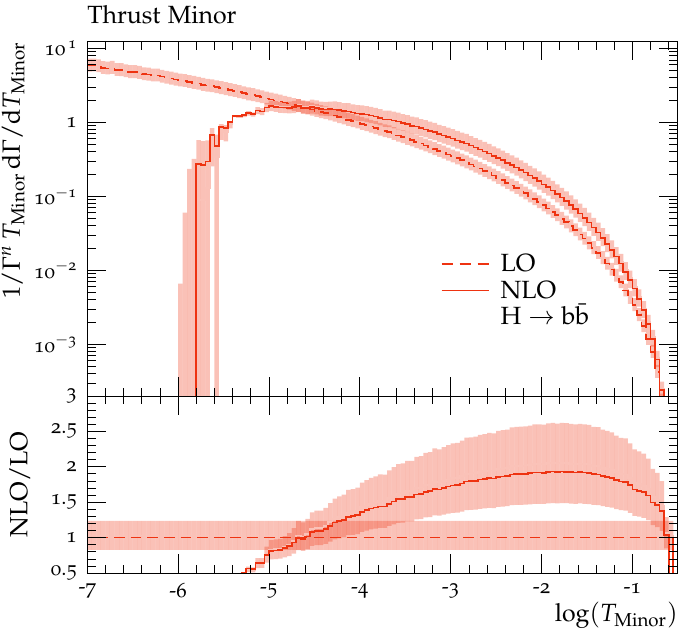}
    \includegraphics[width=0.45\textwidth]{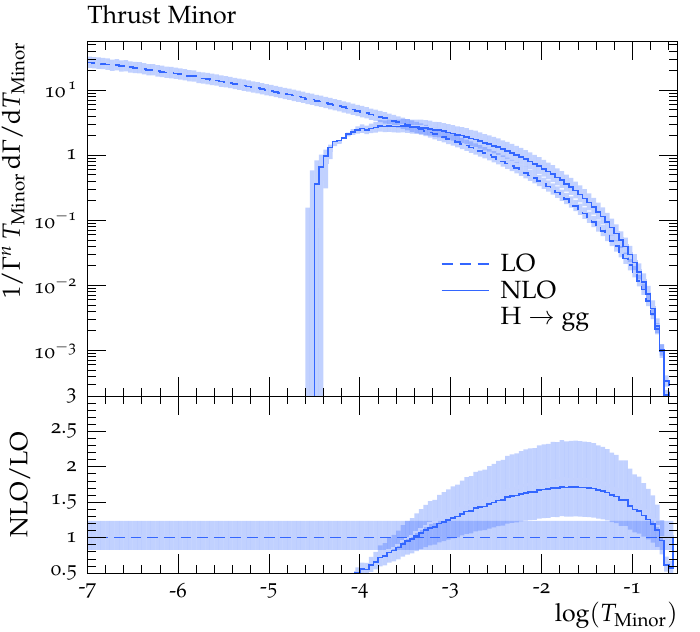} \\
    \caption{Thrust minor in the $\PH\to\Pqb\Paqb$ (\textit{left}) and $\PH\to\Pg\Pg$ (\textit{right}) decay category. The top row shows linearly binned histograms, the bottom row shows the same histograms with logarithmic binning, see main text.}
    \label{fig:TMinor}
\end{figure}

\begin{figure}
    \centering
    \includegraphics[width=0.45\textwidth]{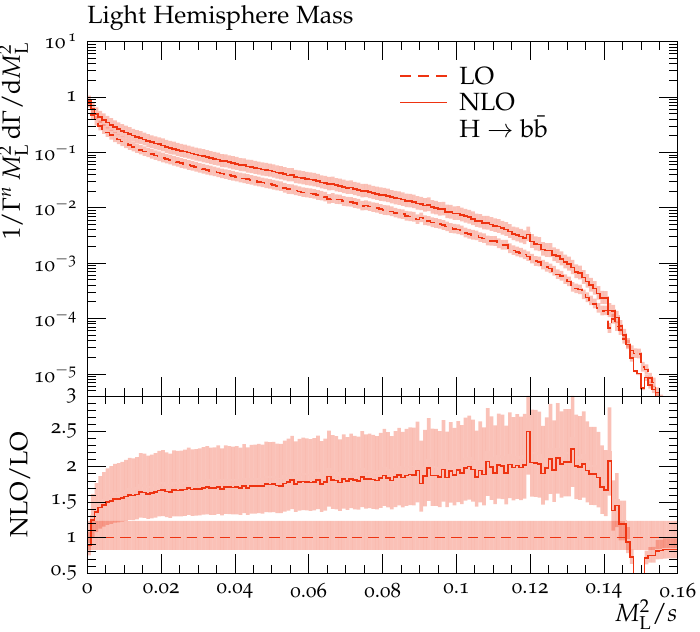}
    \includegraphics[width=0.45\textwidth]{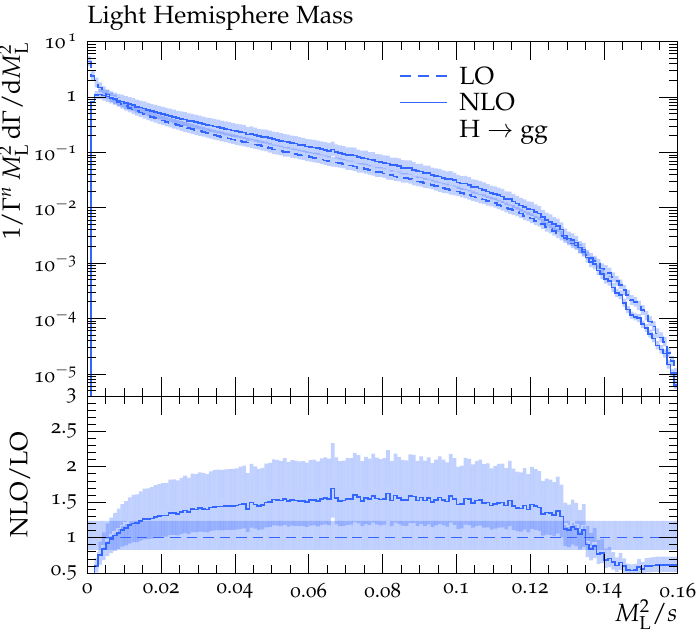} \\
    \includegraphics[width=0.45\textwidth]{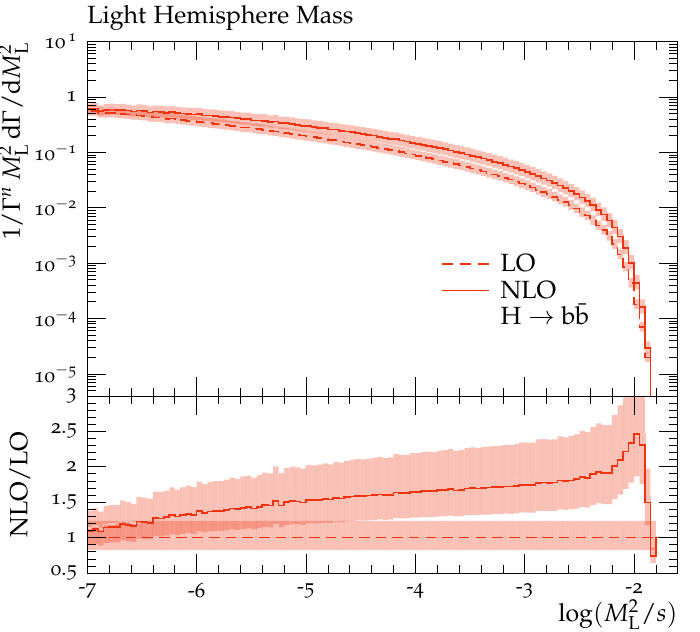}
    \includegraphics[width=0.45\textwidth]{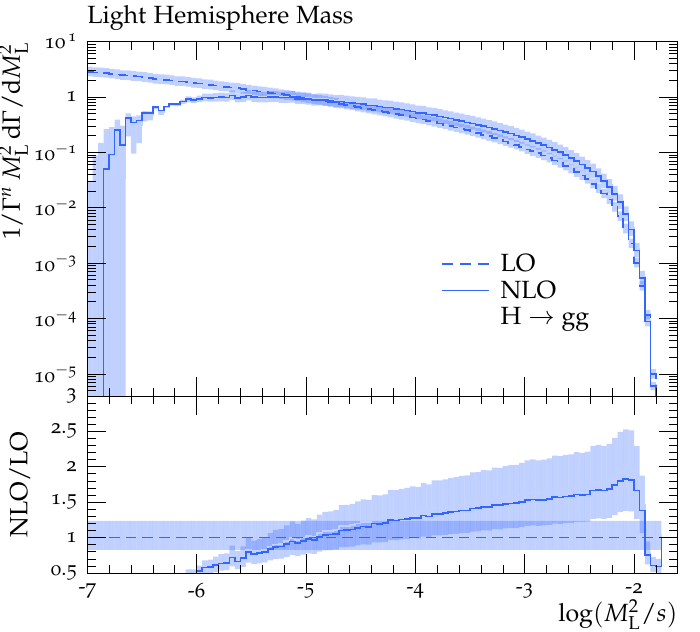}
    \caption{Light hemisphere mass in the $\PH\to\Pqb\Paqb$ (\textit{left}) and $\PH\to\Pg\Pg$ (\textit{right}) decay category. The top row shows linearly binned histograms, the bottom row shows the same histograms with logarithmic binning, see main text.}
    \label{fig:ML2}
\end{figure}

\begin{figure}
    \centering
    \includegraphics[width=0.45\textwidth]{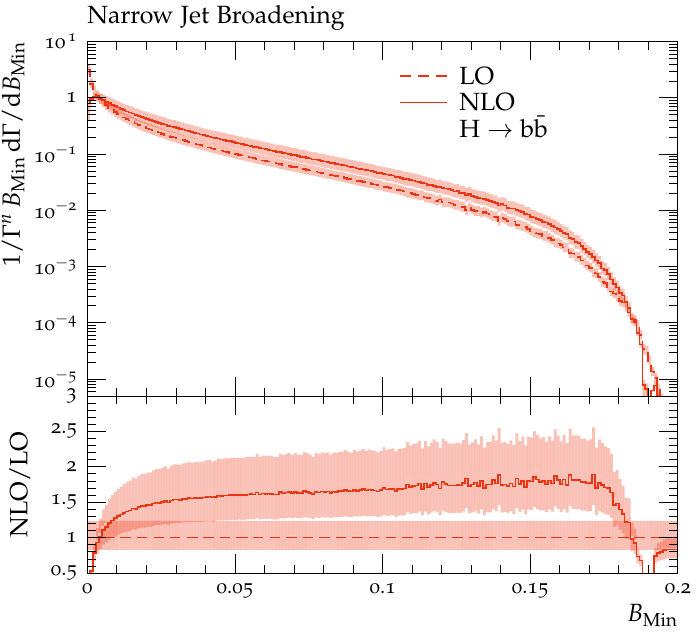}
    \includegraphics[width=0.45\textwidth]{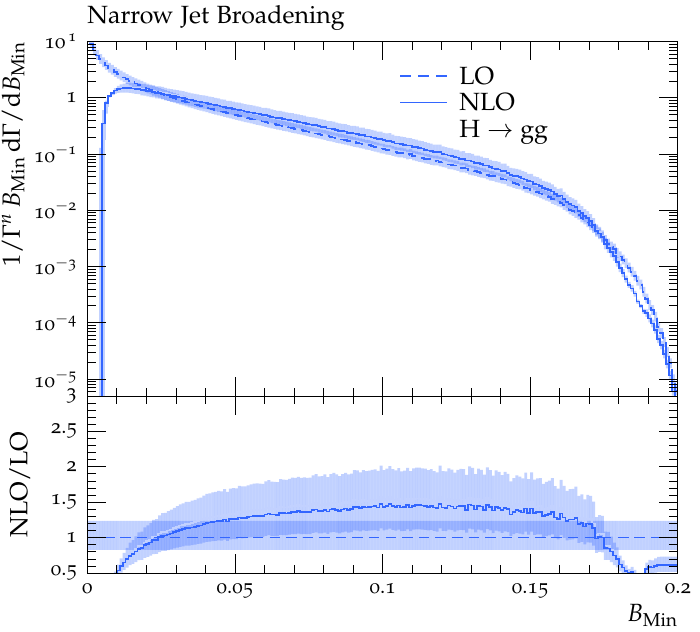} \\
    \includegraphics[width=0.45\textwidth]{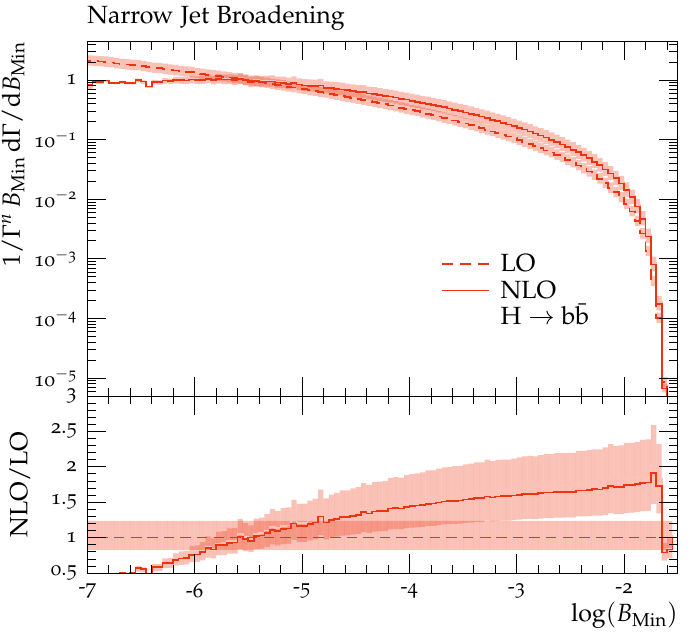}
    \includegraphics[width=0.45\textwidth]{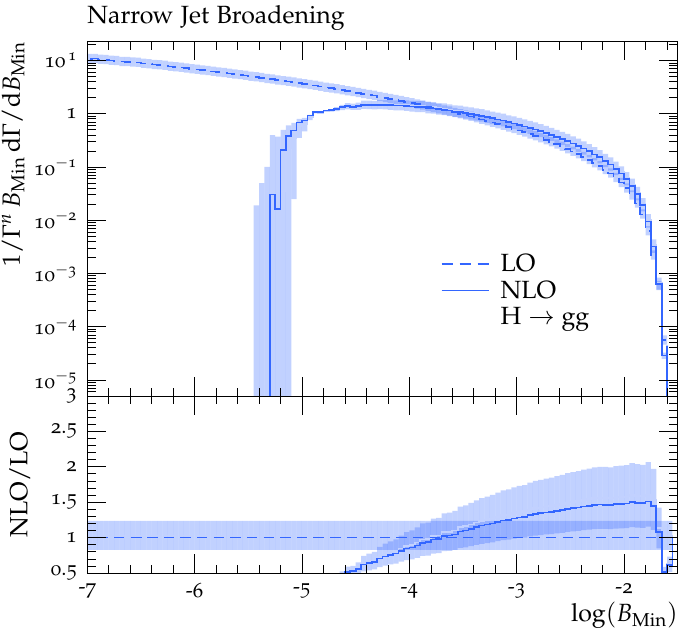} 
    \caption{Narrow jet broadening in the $\PH\to\Pqb\Paqb$ (\textit{left}) and $\PH\to\Pg\Pg$ (\textit{right}) decay category. The top row shows linearly binned histograms, the bottom row shows the same histograms with logarithmic binning, see main text.}
    \label{fig:BMin}
\end{figure}

\begin{figure}
    \centering
    \includegraphics[width=0.45\textwidth]{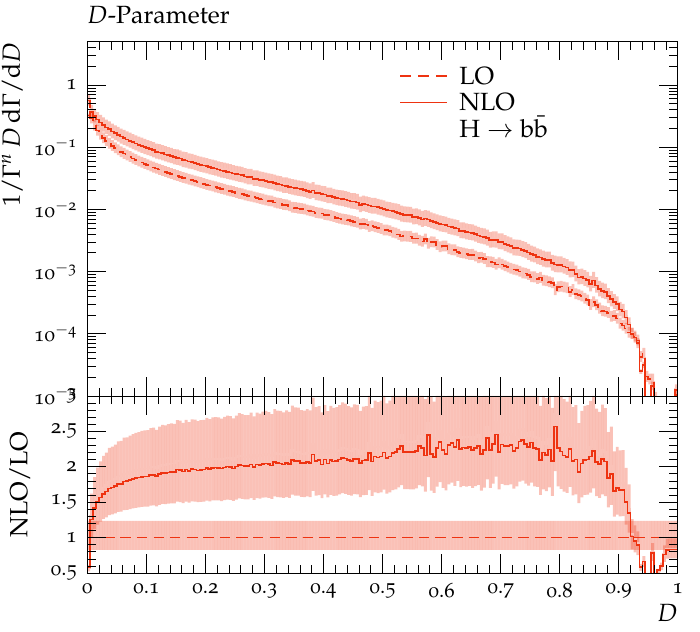}
    \includegraphics[width=0.45\textwidth]{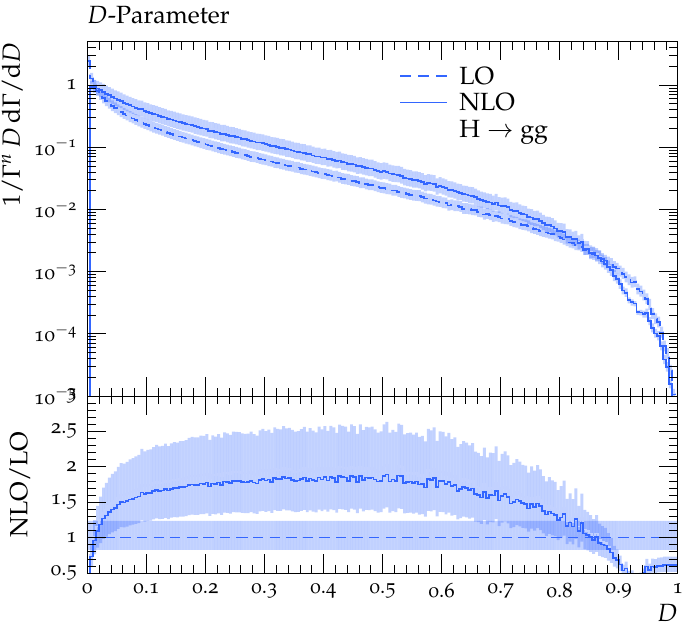} \\
    \includegraphics[width=0.45\textwidth]{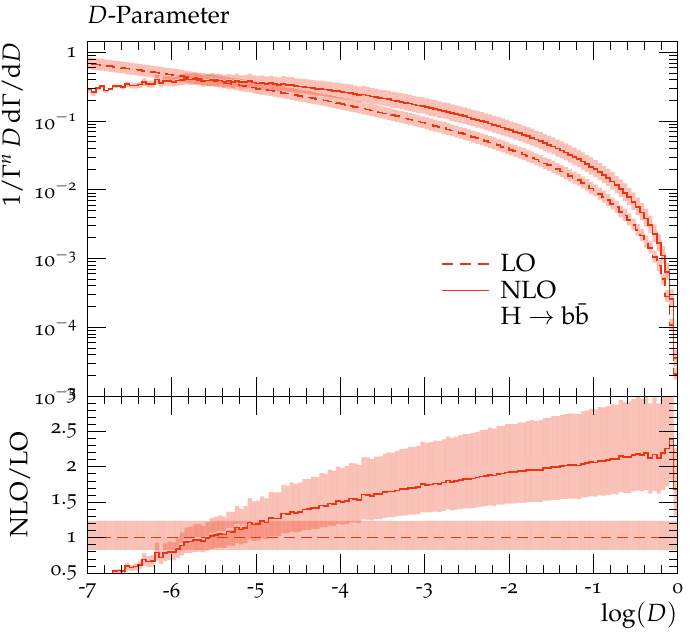}
    \includegraphics[width=0.45\textwidth]{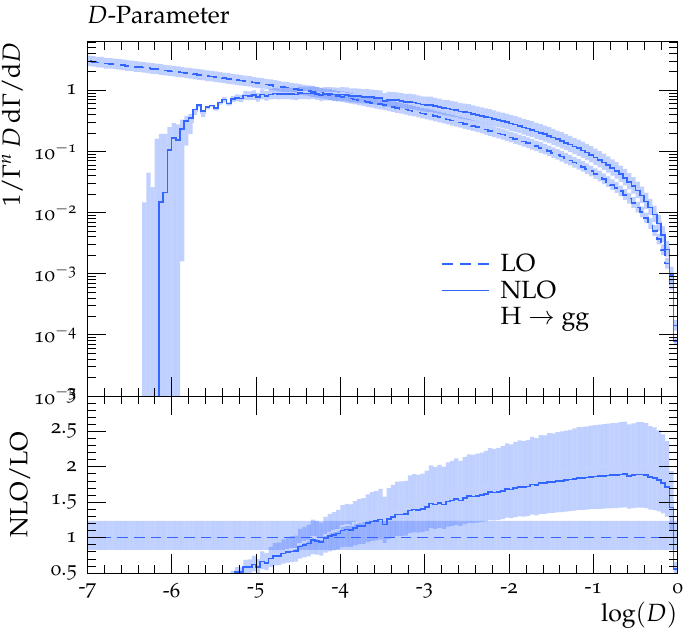}
    \caption{$D$-parameter in the $\PH\to\Pqb\Paqb$ (\textit{left}) and $\PH\to\Pg\Pg$ (\textit{right}) decay category. The top row shows linearly binned histograms, the bottom row shows the same histograms with logarithmic binning, see main text.}
    \label{fig:D}
\end{figure}

\begin{figure}
    \centering
    \includegraphics[width=0.45\textwidth]{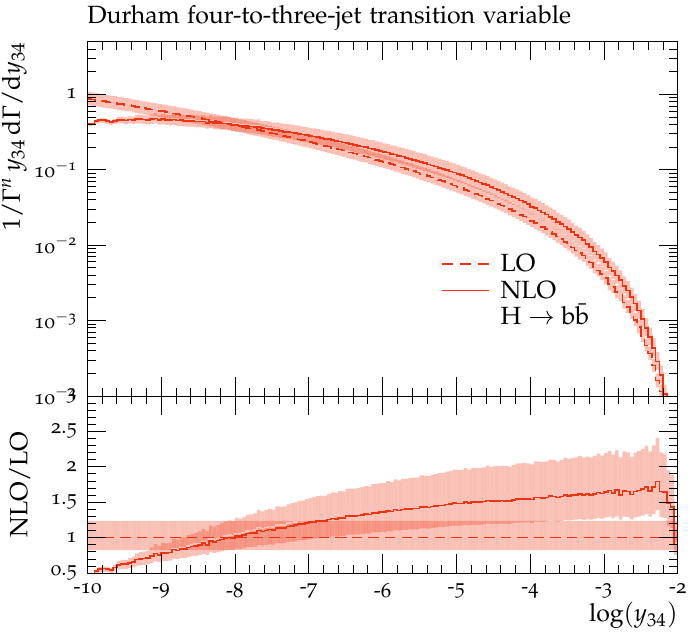}
    \includegraphics[width=0.45\textwidth]{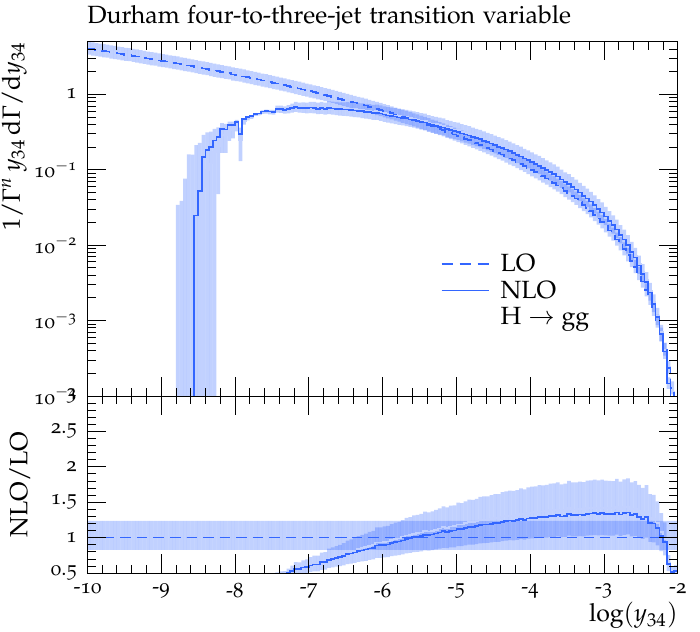}
    \caption{Durham four-to-three jet transition variable denoted by $y_{34}$ in the $\PH\to\Pqb\Paqb$ (\textit{left}) and $\PH\to\Pg\Pg$ (\textit{right}) decay category, see main text.}
    \label{fig:y34}
\end{figure}

Numerical predictions for the five event-shape observables are shown in \cref{fig:TMinor,fig:ML2,fig:BMin,fig:D,fig:y34}. Each figure contains four plots; we present results with linear binning in the top row and results with logarithmic binning in the bottom row; results in the $\PH\to\Pqb\Paqb$ decay category are shown in the left-hand column, while results in the $\PH\to\Pg\Pg$ decay category are shown in the right-hand column. Each plot contains \LO and \NLO predictions, shown with a dashed and solid line, respectively.
Predictions obtained by the scale variation are shown with lighter shading.
The infrared region is located towards the left-hand side of each plot, whereas the hard multi-particle region is located on the right-hand side.
\newline
All distributions exhibit the usual characteristics of event-shape observables at fixed order. The \LO distributions diverge towards positive infinity in the infrared limit on the left-hand side of the plots. In these regions of phase-space, one of the four particles in the \LO Born configuration becomes unresolved and the four-particle configuration assumes the shape of a planar three-jet event. At \NLO, the distributions develop a peak close to zero and diverge towards negative infinity in the infrared limit. As a consequence, all distributions show these characteristic behaviour towards the infrared region. As mentioned before, an accurate description of the observables in this phase-space region requires the resummation of large logarithms, which we do not include in our predictions. Instead as detailed in \cref{subsec:infraredBehaviour}, we restrict ourselves to provide predictions above a minimum value of the observables. 
Generically, we observe rather large \NLO corrections with $K$-factors between $1.3$ and $2.3$ and scale uncertainties of similar size in both categories. It may be surprising that these corrections are numerically slightly bigger for the $\PH\to\Pqb\Paqb$ decay category. The reason for this is an interplay of two effects. On the one hand, predictions in the $\PH\Pg\Pg$ category generally receive larger \NLO corrections, which results in numerically bigger \NLO coefficients $\rd C/\rd O$. On the other hand, the \NLO distribution are normalised by $1/\Gamma^{(1)}$, as opposed to $1/\Gamma^{(0)}$ in the \LO case, cf.~\cref{eq:rate}. Because the \NLO correction to the partial two-particle decay width is again numerically bigger for $\PH\to\Pg\Pg$ decays than for $\PH\to\Pqb\Paqb$ decays ($\sim 1.6$ compared to $\sim 1.2$ at $\muR = M_\PH$), the \NLO distributions are subject to a more sizeable scaling. For the observables considered here, this has the effect that the full \NLO correction becomes numerically smaller in the $\PH\to\Pg\Pg$ decay category.

\paragraph{Thrust minor}
In \cref{fig:TMinor} we present results for the thrust minor at \LO and \NLO in the two decay categories.
In both decay modes, we observe rather large \NLO corrections, with $K$-factors at the central scale around $1.9$ in the $\PH\Pqb\Paqb$ category and $1.7$ in the $\PH\Pg\Pg$ category. In addition, there is a visible shape difference between the two decay modes, as can be inferred from the plots in the top row (with linear binning) of \cref{fig:TMinor}. 
We observe a substantial shift of the peak of the $\PH\Pg\Pg$ distribution away from the infrared region on the left-hand side of the plots. 
Further, the $\PH\Pqb\Paqb$ \NLO correction follows a rather flat shape, while the $\PH\Pg\Pg$ \NLO correction has a more curved shape, as can be seen in the ratio panels of the top-row plots.
This assessment is confirmed by the logarithmically binned distributions shown in the bottom row of \cref{fig:TMinor}. From the ratio panels in the bottom-row plots, it is visible that the intersection of the \LO and \NLO predictions is located at $\log(T_\mathrm{Minor}) = -4.6$ in the $\PH\to\Pqb\Paqb$ case and at $\log(T_\mathrm{Minor}) = -3.4$ in the $\PH\to\Pg\Pg$ case.

\paragraph{Light-hemishere mass}
In \cref{fig:ML2}, \LO and \NLO results for the light-hemisphere mass in the two decay categories are shown. 
From the top row of the figure, with linear binning, it is evident that the peak of the \NLO distributions is located further in the infrared region in both Higgs-decay categories. It cannot be fully resolved in the $\PH\Pqb\Paqb$ category but only in the $\PH\Pg\Pg$ category. This is confirmed by the bottom row of \cref{fig:ML2}, which shows that the intersection of the \LO and \NLO predictions is located at $\log(M_\mathrm{L}^2/s) < -7$ in the $\PH\Pqb\Paqb$ category and at $\log(M_\mathrm{L}^2/s) = -5$ in the $\PH\Pg\Pg$ category. This manifests itself in a shift of the peak of the $\PH\to\Pg\Pg$ distribution towards harder scales with respect to the peak of the $\PH\to\Pqb\Paqb$ distribution, as can be best seen in the plots in the top row of \cref{fig:ML2}.
From the ratio plots, we see that the \NLO corrections are again rather large, with values at the central scale of about $2.0$ in the $\PH\to\Pqb\Paqb$ category and $1.6$ in the $\PH\to\Pg\Pg$ category. Generally, the shape of the \NLO correction is rather similar, being mostly flat for the two decay categories. 

\paragraph{Narrow jet broadening}
\Cref{fig:BMin} shows results for the narrow jet broadening in both decay categories at \LO and \NLO. 
In both decay categories, the peak of the \NLO distributions as well as its shift away from the planar three-jet limit in the $\PH\Pg\Pg$ case is clearly visible in the top-row plots.
The bottom row of \cref{fig:BMin} shows that the intersection of the \LO and \NLO distributions is located at $\log(B_\mathrm{Min}) = -5.5$ in the $\PH\Pqb\Paqb$ decay category, while it is shifted to a significantly larger value of $\log(B_\mathrm{Min}) = -3.7$ in the $\PH\Pg\Pg$ decay category.
The shape of the \NLO corrections at the central scale is rather flat around a factor of $1.8$ for $\PH\Pqb\Paqb$ decays, but slightly curved at around $1.4$ in $\PH\Pg\Pg$ decays. 

\paragraph{$D$-parameter}
In \cref{fig:D} we present \LO and \NLO results for the $D$-parameter in the two decay categories.
We observe large \NLO corrections with $K$-factors at the central scale of $~2.3$ in the $\PH\Pqb\Paqb$ decay and $~1.8$ in the $\PH\Pg\Pg$ decay.
The shape of the \NLO corrections is comparable to thrust minor, as can be seen from comparing the ratio panels in the top row of \cref{fig:D} with the ones in the top row of \cref{fig:TMinor}. Most notably, the shape of \NLO corrections is rather flat for $\PH\Pqb\Paqb$ decays, while it is more curved for $\PH\Pg\Pg$ decays.
The ratio panels in the logarithmically binned distributions in the bottom row of \cref{fig:D} reveal the location of the \LO-\NLO intersection. Specifically, it sits at $\log(D) = -5.6$ for $\PH\Pqb\Paqb$ decays and $\log(D) = -4.2$ for $\PH\Pg\Pg$ decays.

\paragraph{Four-to-three-jet transition variable}
\Cref{fig:y34} contains results for the four-to-three-jet transition variable denoted by $y_{34}$, using the Durham jet algorithm to build jets at \LO and \NLO for the two decay categories. We note that, different to the other event shapes, we plot results for this four-jet resolution distribution only with a logarithmic binning. 
Nevertheless, we observe similar results as before. Events in the $\PH\Pg\Pg$ category are generally clustered earlier and the intersection of the \LO and \NLO curves is shifted by about two orders of magnitude away from the infrared limit, from $\log(y_{34}) = -8.1$ in the $\PH\Pqb\Paqb$ case to $\log(y_{34}) = -5.6$ in the $\PH\Pg\Pg$ case.
The \NLO corrections are rather mild, with $K$-factors at the central scale around $1.7$ for $\PH\Pqb\Paqb$ decays and $1.3$ for $\PH\Pg\Pg$ decays.

\section{Conclusions}
\label{sec:conclusion}
In this paper we have, for the first time, presented results for four-jet-like event-shape observables in hadronic Higgs decays calculated including \NLO corrections in perturbative QCD.
Specifically, we computed the five event shapes related to the four-jet-like event shape variables thrust minor, light-hemisphere mass, narrow jet broadening, $D$-parameter and the four-to-three-jet transition variable in the Durham algorithm. 

Using the antenna subtraction framework, we have implemented our computation in the existing EERAD3 parton-level Monte-Carlo generator extended to deal with hadronic Higgs decay observables, with four-parton final states at Born level. 

The calculation was performed in an effective theory under the assumption of massless $\Pqb$-quarks with finite Yukawa coupling and an infinitely heavy top-quark, leading to the consideration of two distinct class of processes, belonging to the $\PH\Pg\Pg$ and $\PH\Pqb\Paqb$ Higgs-decay categories.

We observe that all of the distributions show the characteristic behaviour of event-shape observables at fixed order in perturbative QCD. Specifically, the \LO predictions diverge towards positive infinity in the infrared limit, whereas the \NLO predictions develop a peak close to the limit before diverging towards negative infinity.

We have shown that all of the event shapes receive large \NLO corrections with $K$-factors at the central scale between $1.7 - 2.3$ in the $\PH\Pqb\Paqb$ decay category and $1.3 - 1.8$ in the $\PH\Pg\Pg$ decay category, depending on the observable. Interestingly, the \NLO corrections are slightly smaller in $\PH\to\Pg\Pg$ decays. This can be explained by the normalisation to the \NLO two-particle decay width, which receives larger corrections in the $\PH\to\Pg\Pg$ case and as such leads to a stronger scaling in this decay category. If, instead, a normalisation to the \LO two-parton decay width was chosen, we would find larger \NLO $K$-factors for distributions in the $\PH\to\Pg\Pg$ category.
In both decay categories, the largest \NLO corrections can be observed for the $D$-parameter and thrust minor. The size of the \NLO corrections for the narrow jet broadening, light-hemisphere mass, and the four-jet resolution is slightly smaller.

Concerning the shape of the distributions inside a given Higgs-decay category, we have shown that all event-shape distributions display very similar behaviour at \LO, whereas the shape visibly changes at \NLO level. For all observables considered here, the shape of the \NLO corrections is rather flat for $\PH\Pqb\Paqb$ decays, while it is more curved in $\PH\Pg\Pg$ decays. 
We further find visible shape differences between the Yukawa-induced and gluonic decay categories at \NLO.
Specifically, we observe a characteristic shift of the peaks of the \NLO distributions away from the infrared limit for $\PH\to\Pg\Pg$ decays, accompanied by a similar shift of the intersection of the \LO and \NLO predictions.

Our calculation provides crucial ingredients for the computation of higher-order QCD corrections for event-shape observables in hadronic Higgs decays.
Most imminently, it gives access to precise predictions including next-to-leading order corrections to four-jet like event-shape observables in all hadronic Higgs-decay channels, as needed for phenomenological studies of Higgs decay properties, at a future lepton collider. While we have focussed only on decays to bottom quarks here, it is straightforward to extend our implementation to decays to other quark flavour species.
We have implemented our computation in such a way that it facilitates the matching to resummed predictions in the future.
Our calculation also provides part of the real-virtual and double-real contributions to the \NNLO calculation of three-jet event-shape observables.
To this end, the \NLO subtraction terms used here need to be suitably extended to \NNLO-type subtraction terms.
As we have here constructed all subtraction terms within the antenna-subtraction framework, this is a straightforward extension and will be subject of future work.

\acknowledgements
The authors are grateful to Stefano Pozzorini for helpful advices on the renormalisation schemes used in OpenLoops2.
The authors also wish to thank Damien Geissbühler for his contribution to the computation of observables in the Hgg category at an early stage of this project.
CTP and AG are supported by the Swiss National Science Foundation (SNF) under contract 200021-197130 and by the Swiss National Supercomputing Centre (CSCS) under project ID ETH5f.
CW is supported by the National Science Foundation through awards NSF-PHY-1652066 and NSF-PHY-201402.

\bibliography{bibliography.bib}

\end{document}